\documentclass[aps,prl,preprint,groupedaddress,showpacs]{revtex4}
\usepackage{epsfig}

\begin{document}     

\title{Transverse Thermoelectric Conductivity of Bilayer Graphene in Quantum Hall Regime } 
\author{Chang-Ran Wang, Wen-Sen Lu, and Wei-Li Lee}      
\affiliation{\textit{Institute of Physics, Academia Sinica, Nankang, Taipei, Taiwan R.O.C.} }

\date{\today}      

\begin{abstract}
We performed electric and thermoelectric transport measurements of bilayer graphene in a magnetic field up to 15 
Tesla. The transverse thermoelectric conductivity $\rm\alpha_{xy}$, determined from four transport coefficients, 
attains a peak value  of $\rm\alpha_{xy, peak}$ whenever chemical potential lies in the center of a Landau level. The 
temperature dependence of $\rm\alpha_{xy, peak}$ is dictated by the disorder width $\rm W_L$. For $\rm 
k_BT/W_L\leq$0.2, $\rm\alpha_{xy, peak}$ is nominally linear in temperature, which gives $\rm\alpha_{xy,peak}/T=0.19 
\pm 0.03 n A/K^2$ independent of the magnetic field, temperature and Landau Level index. At $\rm k_BT/W_L\geq$0.5, 
$\rm\alpha_{xy, peak}$ saturates to a value close to the predicted universal value of $\rm 4\times(ln2)k_Be/h$ 
according to the theory of Girvin and Jonson. We remark that an anomaly is found in $\rm\alpha_{xy}$ near the charge 
neutral point, similar to that in single-layer graphene.            
   
\end{abstract}
\maketitle
Graphene, a single atomic layer of carbon sheet, has attracted considerable attention in the past few years 
\cite{geim}.  Its unusual massless Dirac-fermion excitations, deriving solely from its honeycomb structure, have 
already unveiled many interesting transport phenomena \cite {rtQHE,FQHE,Klein}. More recently, there is a growing 
interest in the bilayer graphene partly due to the observation of tunable band gap by breaking inversion symmetry 
using external bias \cite{bilayer_gap}. In addition, its chiral massive excitations are unique which give rise to the 
Berry phase $\rm 2\pi$ \cite{berry2pi} and the two-fold orbital degeneracy in the zero-energy Landau level (LL). 
Several authors have pointed out that the coulomb exchange interaction within a LL with multi-degeneracy can break the 
symmetry and induce a spin-polarized quantum Hall state \cite{QHFM}. This becomes more pronounced in a bilayer 
graphene in which its zero-energy LL contains eightfold degeneracy (two spin, two valley and two orbital degeneracies) 
\cite{biQHFM}. The lifting of spin degeneracy in the zero-energy LL has been uncovered in single-layer \cite{single} 
and bilayer graphene \cite{bilayer} providing the evidence for exchange enhanced Zeeman splitting.

A fundamental issue in the quantum Hall effect (QHE) is the role of the edge current contribution to the observed 
quantized Hall conductivity in two dimensional electron system (2DES) with finite dimensions \cite{halperin}. A 
current density $\rm\vec{J}$ generated by an electric field $\rm\vec{E}$ and a temperature gradient 
$\rm(-\vec{\bigtriangledown}T)$ can be expressed by 
$\rm\vec{J}=\tensor{\sigma}\vec{E}+\tensor{\alpha}(-\rm\vec{\bigtriangledown}T)$, where $\rm\tensor{\sigma}$ and 
$\rm\tensor{\alpha}$ are the conductivity tensor and thermoelectric conductivity tensor, respectively. By assuming no 
transverse temperature gradient in the steady state, $\rm\tensor\alpha$ can be expressed in terms of conductivity $\rm 
\sigma_{xx}$, Hall conductivity $\rm\sigma_{xy}$, thermopower $\rm S_{xx}$ and Nernst signal $\rm S_{yx}$ as
\begin{equation}
\rm \alpha_{xx}=\sigma_{xx}S_{xx}-\sigma_{xy}S_{yx} and \alpha_{xy}=\sigma_{xy}S_{xx}+\sigma_{xx}S_{yx}.
\label{alphaeq}
\end{equation}
We have used the definition of $\rm S_{xx}\equiv -E_x/|\bigtriangledown T|$ (positive for hole) and $\rm S_{yx}\equiv 
E_y/|\bigtriangledown T|$. Under an intense magnetic field, there exists current-carrying edge states within a 
magnetic length $\rm\ell_B\equiv\sqrt{\hbar/(eB)}$ to the boundary. In the presence of $\rm\vec{E}$, the induced 
difference in the edge currents at opposite sides due to the shift in chemical potential $\rm\mu$ can be calculated to 
give $\rm\delta I_{edge}=\nu e^2V_H/h$, where $\rm\nu$ is an integer and $\rm V_H$ is the Hall voltage. This agrees 
well with the observed quantized Hall conductivity $\rm\sigma_H=\nu e^2/h$ with $\nu$ representing the filling factor. 
An alternative way to generate a non-zero $\rm\delta I_{edge}$ is to apply a $(-\rm\vec{\bigtriangledown}T)$. The 
resulting $\rm\delta I_{edge}$ is also quantized, giving a universal value in the transverse thermoelectric tensor 
$\rm\alpha_{xy}=(ln2)k_Be/h$ whenever $\rm\mu$ sits at the center of a LL according to Girvin and Jonson's (GJ) theory 
\cite{GJ,sheng}. Recently, measurements on single-layer graphene \cite{joe} found that $\rm\alpha_{xy}$ is consistent 
with the predictions of GJ. In this work, we report electric and thermoelectric transport measurements in bilayer 
graphene. $\rm\alpha_{xy}$ is determined from $\rm \sigma_{xx}$, $\rm\sigma_{xy}$, $\rm S_{xx}$ and $\rm S_{yx}$. It 
reaches a peak value of $\rm\alpha_{xy, peak}$ whenever $\rm\mu$ crosses the center of a LL. The temperature and 
magnetic field dependences of $\rm\alpha_{xy, peak}$ are then discussed and compared to the Mott relation \cite{mott}.

\begin{figure}[ht]
\centerline {\epsfig{figure=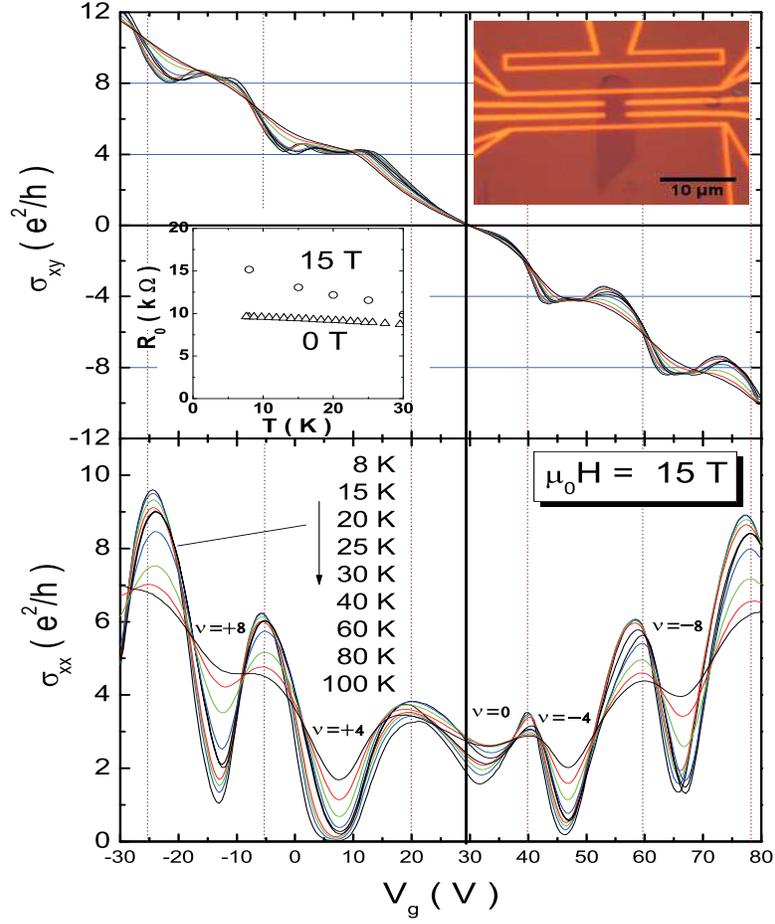,height=4.8in,width=4in,clip=0}}
\caption {\label{QHE} $\rm V_g$ dependence of $\rm\sigma_{xy}$ and $\rm \sigma_{xx}$ at temperatures ranging from 8 K 
to 100 K in S45. An optical image of S45 is shown in the upper-right inset. The middle inset plots the sheet 
resistance at CNP $\rm R_0$ vs. T at 15 Tesla (circle) and 0 Tesla (triangle).} 
\end{figure}

Bilayer graphene samples were prepared by mechanical exfoliation from high quality bulk graphite on a doped silicon 
substrate, which serves as a bottom gate, with 300 nm of silicon dioxide. The number of layer was first identified 
from the optical contrast and then further confirmed by the pattern of $\rm\sigma_{xy}$ plateau in quantum Hall 
regime. The device shown in the inset of Fig. \ref{QHE} was fabricated using electron beam lithography \cite{joe, wei, 
zuev}, comprising four electrodes, a local heater and two local thermometers, followed by Cr(1 nm)/Au(40 nm) film 
deposition and lift-off process. A $(-\rm\vec{\bigtriangledown}T)$ across the bilayer graphene was modulated by 
feeding an alternating current to the local heater at a frequency $\rm\omega\sim$ 3 Hz. The thermoelectric signals 
$\rm S_{xx}$ and $\rm S_{yx}$ are then detected at second harmonic $\rm 2\omega$ using a lock-in amplifier. Similarly, 
the temperature gradient is determined by feeding a direct current to the thermometer while detecting its resistance 
oscillation at $\rm 2\omega$. The $(-\rm\vec{\bigtriangledown}T)$ we used is typically in a range from 10 mK/$\rm\mu 
m$ to 50 mK/$\rm\mu m$. Both $\rm S_{xx}$ and $\rm S_{yx}$ signals are found to scale linearly with 
$(-\rm\vec{\bigtriangledown}T)$ as expected. The results from two devices, labeled as S37 and S45 with different 
sample geometry, show consistent behavior which would be described in detail as follows.

\begin{figure}[ht]
\centerline {\epsfig{figure=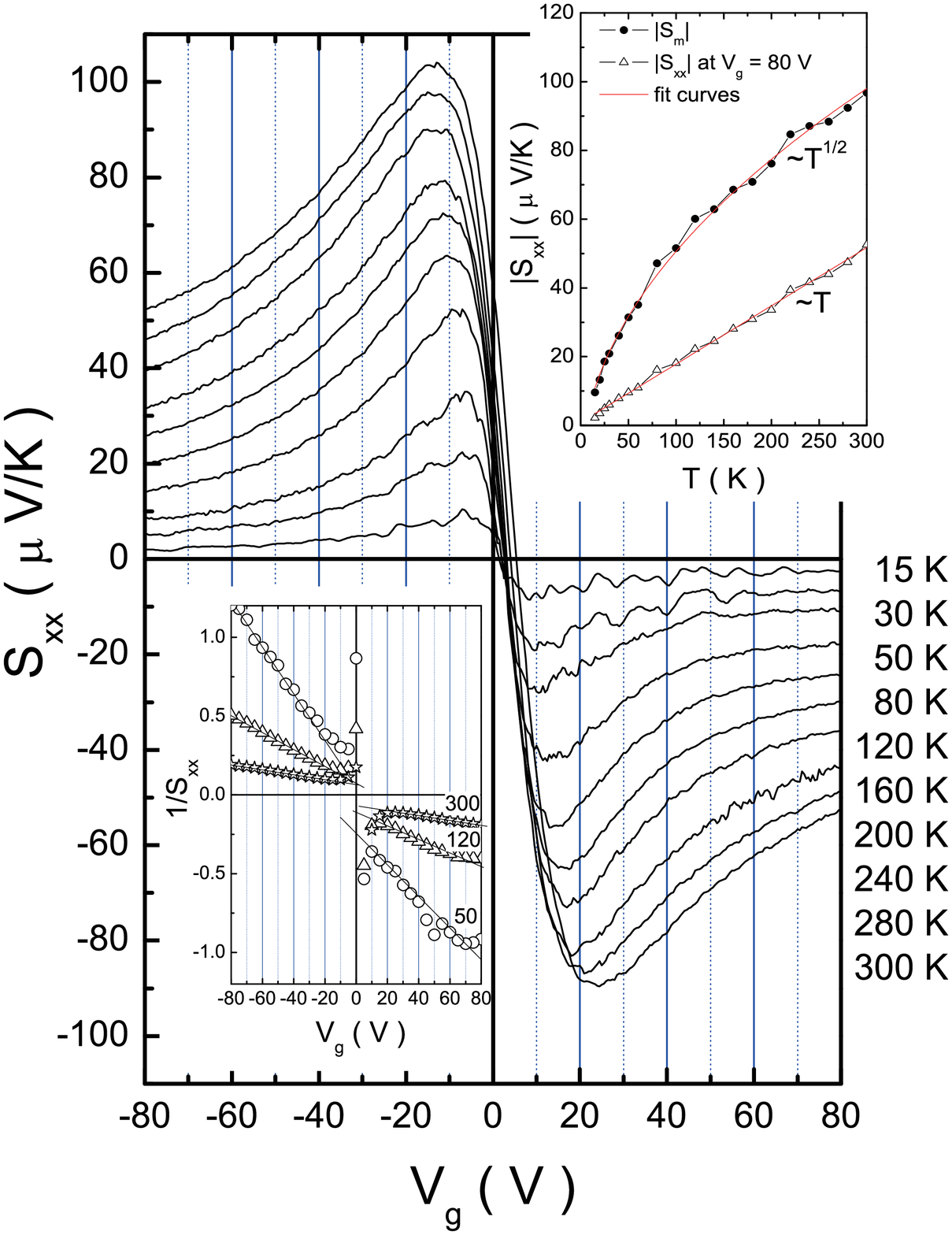,height=4.8in,width=4in,clip=0}}
\caption {\label{ST} Thermopower $\rm S_{xx}$ vs. $\rm V_g$ at temperatures ranging from 15 K to 300 K in S37. The 
upper-right inset shows the T dependence of maximum thermopower $\rm |S_{m}|$ and $\rm |S_{xx}| ( \rm V_g=80 V)$. The 
red lines are the power law fitting to the data points. The lower-left inset plots $\rm 1/\rm S_{xx}$ vs. $\rm V_g$ at 
selected temperatures.} 
\end{figure}

Fig. \ref{ST} shows $\rm S_{xx}$ as a function of applied gate voltage $\rm V_g$ at different temperatures ranging 
from 15 K to 300 K in S37. $\rm S_{xx}$ equals zero at charge neutral point (CNP) ($\rm V_g'\equiv V_g-V_{g0}=0, 
V_{g0}$=+5 V) while its magnitude increases rapidly away from CNP and reaches a maximum value of $\rm S_m$ at $\rm 
|V_g'| \approx$ +15 V corresponding to a carrier density $\rm n_c\approx 3\times 10^{12} cm^{-2}$ determined from its 
Hall coefficient. $\rm S_{xx}$ monotonically increases with increasing temperature giving $\rm |S_m|\approx 95\mu$V/K 
at 300 K. The upper-right inset of Fig. \ref{ST} plots the temperature dependence of $\rm |S_m|$ and $\rm 
|S_{xx}(V_g=80V)|$, in which $\rm T^{1/2}$ and T, respectively, power law fittings shown as red lines agree well with 
the data points. On the other hand, $\rm 1/S_{xx}$ is practically linear with $\rm V_g$ away from CNP ($\rm 
|V_g'|\geq$20 V), which is shown at selected temperature of T=300 K, 120 K and 50 K in the lower-left inset of Fig. 
\ref{ST}. From semi-classical theory for a free electron system, the thermopower can be described by $\rm 
S_{xx}=-\pi^2/2(k_B/e)(k_BT/\epsilon_F)$, where $\rm\epsilon$ is the Fermi energy. Since $\rm\epsilon_F \propto 
k_F^2\propto V_g'$ in a bilayer graphene, the observation of $\rm S_{xx}\propto T/V_g'$ away from CNP shown in the 
inset of Fig. \ref{ST} is then consistent with the free electron model. We also note that the $\rm T^{1/2}$ dependence 
of $\rm S_m$ near CNP is different from the T-linear dependence found in single-layer graphene \cite{joe, wei}. As it 
turns out, the failure of the semi-classical theory near CNP persists to the quantum Hall regime.

As a standard practice, $\rm \sigma_{xx}$($\rm\sigma_{xy}$) and $\rm S_{xx}$($\rm S_{yx}$) under a field were 
(anti)symmetrized with respect to the magnetic field to exclude the contribution from mis-alignment of electrodes. At 
15 Tesla, the observed Hall plateau of $\rm\sigma_{xy}=\nu e^2/h$ in S45 (the upper panel in Fig. \ref{QHE}), where 
$\rm\nu=\pm 4, \pm 8$ is the filling factor, indicates the onset of quantum Hall effect even at temperature up to 100 
K. The $\rm 8 e^2$/h step in $\rm\sigma_{xy}$ across the CNP ($\rm V_{g0}\simeq$+29.4 V) further confirms the bilayer 
graphene nature in S45, where its LL energy, given by $\rm E_n=\hbar\omega_C\sqrt{N(N-1)}$ with the LL index N and 
cyclotron frequency $\rm\omega_C=eB/m^*$, has two-fold orbital degeneracy (N=0 and 1) at zero-energy LL. Using $\rm 
m^*$=0.054$\rm m_e$ \cite{biQHFM}, the LL energy difference $\rm\Delta E=E_2-E_0\simeq$530 K. From low-field Hall 
measurement, we determined the mobility $\rm\mu_c\simeq 2,600 cm^2$/V-sec in S45 and the gate capacitance $\rm 
C_g\simeq$110 aF/$\rm\mu m^2$. If naively using E=($\rm C_g$/e)($\rm\pi\hbar/2m^*)V_g'$, the LL spacing equals 
$\rm\Delta E\simeq$560 K, obtained from $\rm \sigma_{xx}$ peaks in Fig. \ref{QHE}, which is reasonably close to the 
theoretical value (530 K). Similarly, the disorder width $\rm W_L$, defined as the full-width at half-maximum of the 
$\rm \sigma_{xx}$ peaks (lower panel in Fig. \ref{QHE}), gives $\rm\sim$180 K. Therefore, the criteria of $\rm\Delta 
E\gg W_L, k_BT$ for quantum Hall regime is justified. We remark the appearance of double-peak feature in $\rm 
\sigma_{xx}$ with slanted $\rm\sigma_{xy}$ plateau near CNP, which suggests the lifting of spin degeneracy at 
zero-energy LL as pointed out recently by Zhao, {\it et al.} in a bilayer graphene \cite{zhao}. The sheet resistance 
at CNP $\rm R_0$ turns out to be nearly T independent below 30 K in zero field (the middle inset of Fig. \ref{QHE}). 
In contrast, the $\rm R_0$ at 15 Tesla increases by 50\% as T drops to 8 K, which basically agrees with previous 
reports of field-induced insulating state \cite{zhao, feldman}.

\begin{figure}[ht]
\centerline {\epsfig{figure=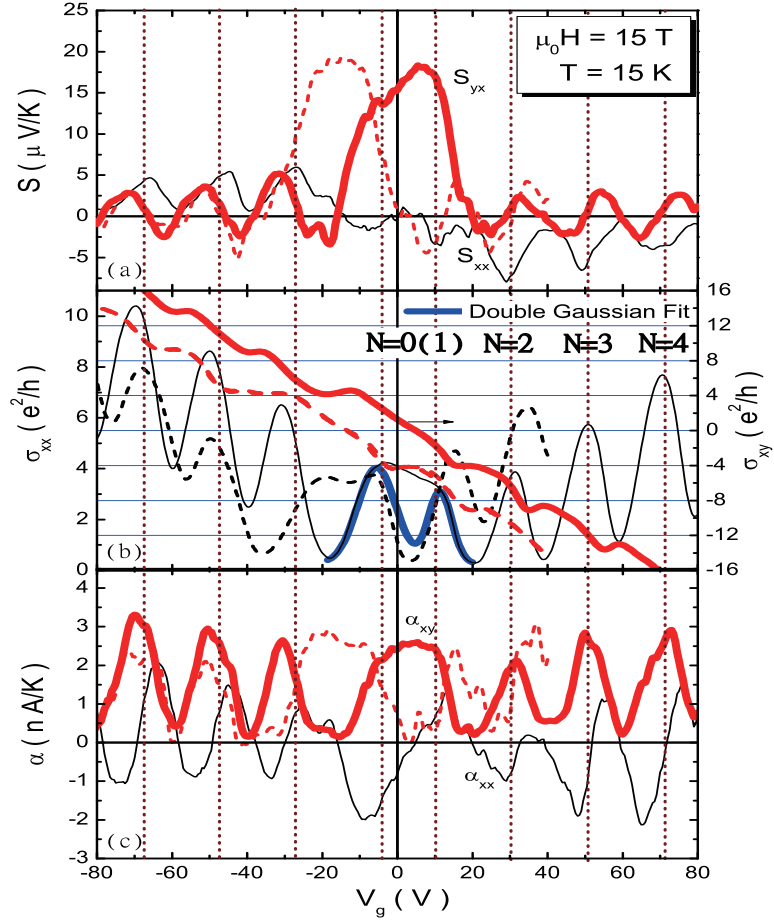,height=4.8in,width=4in,clip=0}}
\caption {\label{aT}$\rm V_g$ dependence of (a) $\rm S_{xx}$ and $\rm S_{yx}$, (b) $\rm \sigma_{xx}$ and 
$\rm\sigma_{xy}$ and (c) $\rm\alpha_{xx}$ and $\rm\alpha_{xy}$ at 15 Tesla and T=15 K in S37. The dashed lines are the 
results from S37 after current annealing. The blue thick line is a double Gaussian fit to $\rm \sigma_{xx}$ near CNP.} 
\end{figure}

Fig. \ref{aT} (a) shows $\rm V_g$ dependence of $\rm S_{xx}$ and $\rm S_{yx}$ in S37 at $\rm\mu_0H$=15 Tesla and T=15 
K shown as thin black line and thick red line, respectively. The corresponding $\rm \sigma_{xx}$ (thin black line) and 
$\rm\sigma_{xy}$ (thick red line) vs. $\rm V_g$ are plotted in Fig. \ref{aT} (b) with N up to 4. A double Gaussian fit 
to the $\rm \sigma_{xx}$ near CNP shown in the thick blue line indicates an excess conductivity near CNP. Both $\rm 
S_{xx}$ and $\rm S_{yx}$ exhibit repeatable oscillations with $\rm V_g$ that effectively shifts the position of 
$\rm\mu$ in bilayer graphene. When $\rm\mu$ crosses the center of a LL, $\rm S_{xx}$ attains a peak value that 
progressively reduces in magnitude at higher N while $\rm S_{yx}$, on the other hand, is nearly zero. Around CNP ($\rm 
V_{g0}\simeq+5$ V), however, a large and broad peak of $\rm S_{yx}\simeq+18\mu$ V/K is observed instead. Using Eq. 
\ref{alphaeq}, the obtained $\rm\alpha_{xx}$ and $\rm\alpha_{xy}$ as a function of $\rm V_g$ are shown in Fig. 
\ref{aT} (c). $\rm\alpha_{xy}$ is positive and reaches a local maximum value of $\rm\alpha_{xy, peak} \simeq$ 3 n A/K 
for N=0(1), 2, 3 and 4 when $\rm\mu$ crosses the center of a LL, while $\rm\alpha_{xx}$ approaches zero instead. We 
also performed a current annealing ($\rm j\sim 3\times 10^8 A/cm^2$) on S37 , which causes the shift of $\rm V_{g0}$ 
from +5 V to - 13 V as demonstrated by dashed lines in Fig. \ref{aT}. Nevertheless, no other major variations in $\rm 
V_g$ dependence are found to result from the current annealing process.

\begin{figure}[ht]
\centerline {\epsfig{figure=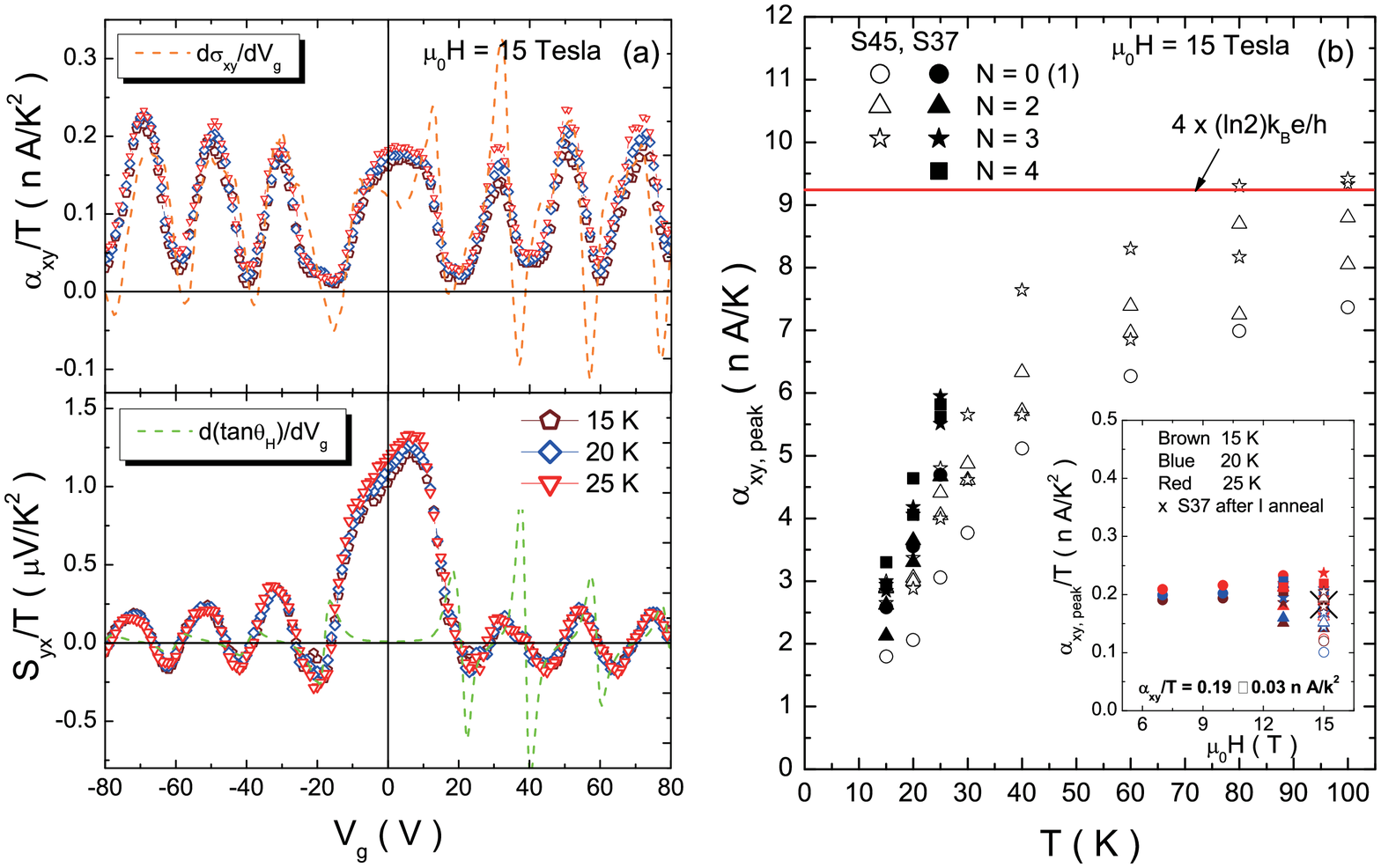,height=4in,width=6in,clip=0}}
\caption {\label{axyoT} (a) $\rm V_g$ dependence of $\rm\alpha_{xy}/T$ (upper panel) and $\rm S_{yx}/T$ (lower panel) 
in S37 at 15 Tesla and three temperatures T=15 K (pentagon), 20 K (diamond) and 25 K (triangle). The orange (upper) 
and green (lower) dashed lines represent the Mott relation fits from $\rm d\sigma_{xy}/dV_g$ and $\rm 
d(tan\theta_H)/dV_g$, respectively. (b) shows the T dependence of $\rm\alpha_{xy, peak}$ in S37 (closed symbols) and 
S45 (open symbols) at 15 Tesla and different LLs N=0(1) (circle), 2 (triangle), 3 (star) and 4 (square). The 
lower-right inset plots the $\rm\alpha_{xy, peak}/T$ vs. $\rm\mu_0H$ with data points sharing the same symbols as 
(b).} 
\end{figure}

In order to further explore the temperature dependence of $\rm\alpha_{xy}$, we performed measurements at different 
temperatures up to 100 K. At low temperatures and 15 Tesla, both $\rm\alpha_{xy}$ and $\rm S_{yx}$ are practically 
linear in T as demonstrated in Fig.\ref{axyoT} (a), where $\rm\alpha_{xy}$/T-$\rm V_g$ at T=15 K, 20 K and 25 K all 
collapse onto a single curve, and so does $\rm S_{yx}$/T-$\rm V_g$. According to semi-classical theory (generalized 
Mott relation), $\rm\alpha_{xy}$ and $\rm S_{yx}$ can be described by $\rm 
\alpha_{xy}=\frac{\pi^2}{3}\frac{k_B^2}{e}T(\frac{\partial\sigma_{xy}}{\partial\varepsilon})_{\epsilon=\mu}$ and $\rm 
S_{yx}=\frac{\pi^2}{3}\frac{k_B^2}{e}T(\frac{\partial tan\theta_{H}}{\partial\varepsilon})_{\epsilon=\mu}$, where $\rm 
tan\theta_H\equiv\sigma_{xy}/\sigma_{xx}$ is the Hall angle. Therefore, we expect that $\rm\alpha_{xy}/T \propto 
(\frac{\partial\sigma_{xy}}{\partial V_g})(\frac{\partial V_g}{\partial\varepsilon})_{\epsilon=\mu}$. Similarly, $\rm 
S_{yx}/T \propto (\frac{\partial tan\theta_H}{\partial V_g})(\frac{\partial 
V_g}{\partial\varepsilon})_{\epsilon=\mu}$. The calculated $\rm\frac{\partial\sigma_{xy}}{\partial V_g}$ and 
$\rm\frac{\partial tan\theta_{H}}{\partial V_g}$ are shown as orange and green dashed lines, respectively, in Fig. 
\ref{axyoT} with arbitrary units, which qualitatively agree with the measured $\rm\alpha_{xy}$/T and $\rm S_{yx}$/T 
except near the CNP. The term $\rm(\frac{\partial V_g}{\partial\varepsilon})_{\epsilon=\mu}$ is proportional to the 
density of state that gives minor influence to the behavior. We noticed the apparent electron-hole asymmetry in the 
magnitude of the calculated $\rm\frac{\partial\sigma_{xy}}{\partial V_g}$ and $\rm\frac{\partial 
tan\theta_{H}}{\partial V_g}$, which may result from the shorting of voltage leads to the ``hot-spot" near the current 
leads \cite{hot}. On the contrary, $\rm S_{xx}$ and $\rm S_{yx}$, generated by a $\rm(-\vec\nabla T)$ across bilayer 
graphene, exhibit much less asymmetry due to the hot-spot shorting effect. The deviation from the calculated value is 
evident near CNP and more dramatic in $\rm S_{yx}$. The temperature dependence of $\rm\alpha_{xy, peak}$ in S37 and 
S45 at 15 Tesla and N=0(1), 2, 3 and 4 are shown in Fig. \ref{axyoT} (b), where closed symbols and open symbols refer 
to the values in S37 and S45, respectively. Even though there is a minor variation in $\rm\alpha_{xy, peak}$ at 
different Ns and different samples, it basically follows T linear dependence below 30 K ($\rm k_BT/W_L\simeq$0.2). The 
corresponding values of $\rm\alpha_{xy, peak}/T$ are plotted as a function of $\rm\mu_0 H$ in the inset of Fig. 
\ref{axyoT} (b), averaged to a constant value of $\rm\alpha_{xy, peak}/T=0.19 \pm 0.03 nA/K^2$ independent of 
$\rm\mu_0H$, N, and T. As T increases above 80 K ($\rm k_BT/W_L\simeq$ 0.5), it flattens up to a nearly constant value 
of $\rm\alpha_{xy, peak}\simeq 8.5\pm$1 nA/K, close to the universal value of $\rm 4\times(ln2)k_Be/h$ (red thick line 
in Fig. \ref{axyoT}).

The disorder width $\rm W_L$, estimated to be $\sim$150 K and $\sim$180 K in S37 and S45, respectively, turns out to 
be an important parameter for the behavior of $\rm\alpha_{xy}$ shown in Fig. \ref{axyoT} (b). At low temperature $\rm 
k_BT \ll W_L$, the number of disorder-induced extended states participating the diffusive transport increases with T, 
which gives rise to the T linear dependence in $\rm\alpha_{xy, peak}$ similar to that in a ferromagnetic metal 
\cite{wlee}. When $\rm k_B$T raises to a value comparable to $\rm W_L$, further increment in T no longer includes more 
extended states. Therefore, $\rm \rm\alpha_{xy, peak}$ approaches the universal value of $\rm 4\times(ln2)k_Be/h$. We 
do, however, observe a smaller $\rm\alpha_{xy,peak}$ value ($\rm\sim$18\% lower) at N=0(1) near CNP. The broad peak of 
$\rm\alpha_{xy}$ at N=0(1) is in big contrast to the sharp and narrow peak observed in a single-layer graphene at 
similar field strength \cite{joe}. We attribute this to be the lifting of spin degeneracy in the bilayer graphene, 
inferred from the double peak feature in $\rm \sigma_{xx}$ that is attainable at lower field due to higher degeneracy 
at zero-energy LL \cite{biQHFM}. A crude estimation of the spin splitting energy $\rm \bigtriangleup E_{spin}$ from 
the double-peak in $\rm \sigma_{xx}$ at 15 Tesla gives $\rm\bigtriangleup E_{spin}\simeq$26 meV ($\sim$300 K) which is 
15-fold larger than the regular Zeeman energy at the same field. In addition, the failure of Mott relation near CNP 
observed in bilayer graphene (Fig. \ref{axyoT} (a)) may result from the proposed novel phase of counter-propagating 
edge channels with opposite spin in spin-polarized quantum Hall regime \cite{QHFM}. We also remark that the behavior 
of $\rm\alpha_{xy}$ is well reproduced in S37 and S45 with sample width to length aspect ratio of 1.3 and 2.4, 
respectively. Current annealing was found to cause insignificant influence except shifting the $\rm V_{g0}$. The 
anomaly near CNP, arising either from an intrinsic effect of the chiral fermion or from an extrinsic effect such as 
electron-hole puddles \cite{Das Sarma}, remains an open question. Further investigation in samples with higher 
mobility is required in order to resolve this issue.

In summary, we performed electric and thermoelectric transport measurements of bilayer graphene in quantum Hall 
regime. Double peak feature in $\rm \sigma_{xx}$ at N=0(1) LL suggests the lifting of the spin degeneracy due to the 
possible exchange-enhanced Zeeman coupling. The disorder width $\rm W_L$ separates two different regimes for the 
behavior of $\rm\alpha_{xy, peak}$. For $\rm k_BT/W_L\leq$0.2, $\rm\alpha_{xy, peak}$ is practically linear in T, 
giving $\rm \alpha_{xy,peak}/T=0.19\pm 0.03 nA/K^2$ independent of the magnetic field, temperature and LL index N. For 
$\rm k_BT/W_L\geq$0.5, $\rm\alpha_{xy, peak}$ saturates to a value of $\rm\alpha_{xy, peak}\simeq 8.5\pm$1 nA/K close 
to the predicted value of $\rm 4\times(ln2)k_Be/h\simeq 9.24$ nA/K based on GJ theory of edge current model. We also 
found anomalous behaviors in $\rm S_{yx}$ and $\rm\alpha_{xy}$ near CNP, where semi-classical theory does not provide 
satisfactory explanation. This may imply the existence of a novel phase of counter-propagating edge channels with 
opposite spin in spin-polarized quantum Hall regime, which may have potential application in spin-electronics.

The authors acknowledge the funding support from Nation Science Council in Taiwan and technical support from the Core 
Facilities for Nanoscience and Nanotechnology at Academia Sinica in Taiwan.

Note added.- During the preparation of this manuscript, we became aware of a related work at lower field ($\rm\leq$ 7 
Tesla) by Nam et al. \cite{nam}, which shows consistent result with our data.    

\clearpage

\end{document}